\documentclass[]{emulateapj}%
\usepackage{apjfonts}
\usepackage{url}
\usepackage{bm}
\usepackage{amsmath}
\usepackage{times}
\slugcomment{Version: \today}
\newcommand\jcap{{J. Cosmology Astropart. Phys.}}

\begin{document}
\title{Probing the Cosmic X-ray and MeV Gamma-ray Background Radiation through the Anisotropy} 

\author{Yoshiyuki Inoue\altaffilmark{1}, Kohta Murase\altaffilmark{2}, Grzegorz M. Madejski\altaffilmark{1}, \& Yasunobu Uchiyama\altaffilmark{1,3}} 

\affil{$^1$Kavli Institute for Particle Astrophysics and Cosmology, Department of Physics, Stanford University and SLAC National Accelerator Laboratory, 2575 Sand Hill Road, Menlo Park, CA 94025, USA}
\affil{$^2$Hubble Fellow, School of Natural Sciences, Institute for Advanced Study, 1 Einstein Dr. Princeton NJ 08540}
\affil{$^3$Department of Physics, Rikkyo University, 3-34-1 Nishi-Ikebukuro, Toshima-ku, Tokyo, Japan 171-8501}
\email{yinoue@slac.stanford.edu}

\begin{abstract}
While the cosmic soft X-ray background is very likely to originate from individual Seyfert galaxies, the origin of the cosmic hard X-ray and MeV gamma-ray background is not fully understood. It is expected that Seyferts including Compton thick population may explain the cosmic hard X-ray background. At MeV energy range, Seyferts having non-thermal electrons in coronae above accretion disks or MeV blazars may explain the background radiation. We propose that future measurements of the angular power spectra of anisotropy of the cosmic X-ray and MeV gamma-ray backgrounds will be key to deciphering these backgrounds and the evolution of active galactic nuclei (AGNs).  As AGNs trace the cosmic large-scale structure, spatial clustering of AGNs exists. We show that {\it e-ROSITA} will clearly detect the correlation signal of unresolved Seyferts at 0.5--2 keV and 2--10 keV bands and will be able to measure the bias parameter of AGNs at both bands. Once the future hard X-ray all sky satellites achieve the sensitivity better than $10^{-12}\ {\rm erg/cm^2/s}$ at 10--30 keV or 30--50 keV - although this is beyond the sensitivities of current hard X-ray all sky monitors - angular power spectra will allow us to independently investigate the fraction of Compton-thick AGNs in all Seyferts. We also find that the expected angular power spectra of Seyferts and blazars  in the MeV range are different by about an order of magnitude, where the Poisson term, so-called shot noise, is dominant. Current and future MeV instruments will clearly disentangle the origin of the MeV gamma-ray background through the angular power spectrum. 
\end{abstract}

\keywords{cosmology: diffuse radiation --  galaxies: active -- X-rays: diffuse background -- gamma rays : theory}

\section{Introduction}
\label{sec:intro}

The cosmic X-ray background (CXB) is an isotropic, apparently diffuse X-ray emission in the Universe which was discovered about 50 years ago \citep{gia62}. It is often assumed that the CXB has been conclusively shown to be the integrated light produced via the accretion process of active galactic nuclei (AGNs), in particular Seyferts, hosting supermassive black holes. This might be correct below $\sim$ 5 keV. Emission from active galaxies has indeed been resolved by the deep X-ray surveys by {\it Chandra} in the broad 0.5--2 keV and 2--10 keV bands. Those objects account for 80--90\% of the CXB \citep{mus00,gia02,ale03a,ale03b,bau04}. However, energy-resolved studies indicate that the resolved fraction of the CXB decreases with energy as 80--90\% over 2--8 keV, $\sim60$\% over 6--8 keV, and $\sim50$\% beyond 8 keV \citep{wor04, wor05}. 

Above $\sim2$ keV, the CXB cannot be due to superposition of unabsorbed AGNs, mainly type I Seyferts.  Those objects show a typical continuum photon index of $\Gamma=1.9$ below 10 keV \citep{nan94,ree00,pic05}, different from that of CXB $\Gamma =1.4$ at 2--8 keV \citep{del04}.  Instead, this unresolved, hard component is generally attributed to the emission from absorbed Seyferts, the so-called type II Seyferts, which might be buried in dusty tori.  A superposition of such sources with varying degrees of photoelectric absorption by the circumnuclear material can cause the total spectrum to appear harder than spectra of unabsorbed Seyferts, but this requires some fine-tuning of absorption properties of sources as a function of redshift and luminosity.  Various population synthesis models successfully explain the CXB by introducing appropriate number of absorbed Seyferts \cite[see e.g.][]{ued03,tre05,gil07}. However, recent studies \citep{tre09} showed that the number of Compton thick AGNs, which are a class of absorbed Seyferts and whose column density is larger than the inverse of the Thomson cross section, is a factor of 3--4 less than that expected in the population synthesis models at least locally \citep[see also][]{aje12}. This may pose a serious problem to our current knowledge of the origin of the CXB.

By contrast, the origin of the cosmic MeV gamma-ray background at $\sim1-10$ MeV has been an intriguing mystery. The Seyfert spectra adopted in population synthesis models of the CXB cannot explain this component because of the assumed exponential cutoff at a few hundred keV, where thermal hot corona above the accretion disk is assumed. Above 100 MeV, it is known that superposition of blazars \citep[e.g.][]{pad93,ino09,abd10,aje12_fermi}, starburst galaxies \citep[e.g.][]{sol99_gal,ack12_stb}, and radio galaxies \citep[e.g.][]{pad93,ino11,dim13} explains most of the total background flux. These populations may contribute to the MeV background as well.
However, the background spectrum from several hundreds keV to several tens MeV is smoothly connected to the CXB spectrum and is much softer (photon index $\Gamma\sim$ 2.8) than the GeV component \citep{fuk75,wat97,wei00}, indicating a different origin from that above 100 MeV \citep[e.g.][]{sre98,abd10_egrb}. 

A few candidates have been proposed to explain the MeV background. One was the nuclear-decay gamma-rays from Type Ia supernovae \citep[SNe Ia;][]{cla75,zdz96,wat99}. However, on the basis of the latest measurements of the cosmic SN Ia rates, recent studies show that the MeV background flux expected from SNe Ia is about an order of magnitude lower than observed \citep{ahn05_sn,str05,hor10}. Seyferts can naturally explain the MeV background including the smooth connection to the CXB \citep{sch78,fie93,ino08}. Comptonized photons produced by non-thermal electrons in hot coronae surrounding accretion disks can produce the MeV power-law tail \citep{ino08}. Such non-thermal electrons are expected to exist if the corona is heated by magnetic reconnection \citep{liu02}. There is also a class of blazars, called MeV blazars, whose spectra peak at MeV energies \citep{blo95,sam06}. These MeV blazars could potentially contribute to the MeV background as well \citep{aje09}. Radio galaxies have been also discussed as the origin of the MeV background \citep{str76}. However, recent studies show that the expected background flux from radio galaxies is $\sim10$\% of the total MeV background flux \citep{mas11,ino11}. Annihilation of the dark matter particles has also been discussed \citep{oli85,ahn05_dm1,ahn05_dm2,and06,ras06,law08}, but those are less natural dark matter candidates, with a mass scale of MeV energies, rather than GeV-TeV dark matter candidates. In either case, there is little observational evidence of MeV emission from these candidates and a quantitative estimate is not easy due to the sensitivity of the MeV measurements.

The angular power spectrum of the background radiation will shed new light on these problems, since it reflects the distribution of its origin in the entire sky. The angular power spectrum is obtained by performing a spherical harmonics transformation of the sky intensity map after subtracting foregrounds and point sources. As an aside, both theoretical and observational studies of the cosmic microwave background (CMB) anisotropy has allowed us to precisely determine the total content in the Universe \citep[e.g.][]{kom09,kom11}. In the gamma-ray sky, the anisotropy is becoming key to understanding the origin of the GeV background \citep{and06,and07_blz,and07_dm,ack12_ani,cuo12,har12,and13}. 

The anisotropy in the X-ray band has been well studied with tools such as auto-correlation functions \citep[e.g.][]{dez90,car93,che94,sol94,sol99,sch00,sol01,sli01,kus02} and cross-correlations with galaxies, clusters, and CMB \citep[e.g.][]{lah93,miy94,car95,bar95,roc95,sol96,tre96,sol97,new99,ste02,bou04_cmb,bou05}. Theoretically, analytical formalism has been developed to calculate the angular power spectra of the cosmic background radiation in X-ray and gamma-ray band \citep[e.g.][]{gao90,lah97,bar98,and06,and07_blz,and07_dm}. However, angular power spectra of the cosmic background radiation from Seyferts and blazars have not been studied in the context of the latest X-ray luminosity function (XLF) in the X-ray and MeV gamma-ray range, while those from SNe Ia \citep{zha04} and MeV dark matter \citep{and06} have been discussed extensively. 

Galaxies and AGNs are hosted by dark matter halos, but they trace the dark matter distribution with some bias. This bias factor is a key to understanding  the formation mechanism, environment, and evolution of AGNs, since it represents the clustering strength of a source population compared with dark matter. The bias parameter determined from various AGN surveys is controversial. While the correlation functions of the X-ray local AGNs detected by the ROentgen SATellite ({\it ROSAT}) suggest the value close to unity \citep{vik95,mul04_bias}, those from {\it Chandra} and X-ray Multi--Mirror Mission -- Newton ({\it XMM-Newton}) suggest stronger clustering \citep{yan03,bas05,gan06}.  Moreover, the bias parameter inferred from X-ray AGNs is higher at $0<z<3$ that from optically selected quasars \citep{kou13}.

Angular power spectrum allows us to study the bias factor of AGNs from another aspect, since the correlation term of angular power spectrum depends on the bias parameter. The {\it Ginga} satellite has studied the angular structure of the CXB down to 0.2 degrees at 4--12 keV in the regions of the North Galactic Pole and the North Ecliptic Pole \citep{car93}. Since no significant deviation from isotropy is found, the bias parameter of AGNs cannot be constrained. The previous analysis of the High Energy Astronomy Observatory ({\it HEAO})1 A--2 X-ray sky map with a XLF of Seyferts indicated the bias parameter close to unity \citep{bou04}, whereas \citet{sch00} showed the angular power spectrum of the {\it HEAO}1 A--2 X-ray sky map is dominated by the shot noise which is independent of the bias parameter. \citet{rev08} reported CXB intensity variation up to $\sim2$\% on angular scales of 20--40 degrees with the Rossi X-ray Timing Explorer (RXTE). However, fluctuation at the smaller angular scales or the detection of the correlation term was not reported.
 
Here, we evaluate the angular power spectra in the soft X-ray region, the hard X-ray region, and the MeV region with the latest Seyfert and blazar XLFs. The new era of the X-ray and MeV gamma-ray Universe is nearing, with current and future missions such as {\it Astro-H} \citep{tak12}\footnote{\url{http://astro-h.isas.jaxa.jp}}, Nuclear Spectroscopic telescope array \citep[{\it NuStar};][]{har10}\footnote{\url{http://www.nustar.caltech.edu}}, extended ROentgen Survey with an Imaging Telescope Array  \citep[{\it e-Rosita};][]{mer12}\footnote{\url{http://www.mpe.mpg.de/eROSITA}}, CAST \citep{nak12}, DUAL \citep{bal12}, GRIPS \citep{gre12}\footnote{\url{http://www.grips-mission.eu}}, and SMILE \citep{tak11} \footnote{\url{http://www-cr.scphys.kyoto-u.ac.jp/research/MeV-gamma/en/}}. Here, we discuss the detectability of the anisotropy at each energy region by these future missions and future possible studies through the anisotropy.

This paper is organized as follows. In Section 2, we present the spectral model of Seyferts and blazars. In Section 3, we describe the XLF of Seyferts and blazars. In Section 4, we briefly review the formulations to calculate the angular power spectra of X-ray and MeV background anisotropy. Results of the angular power spectra are shown in Section 5. Discussions and conclusions are given in Section 6. Throughout this paper, we adopt the standard cosmological parameters of $(h, \Omega_M, \Omega_\Lambda) = (0.7, 0.3, 0.7)$.

\section{X-ray and MeV Gamma-ray Emission from Active Galactic Nuclei}
\label{sec:sed}
\subsection{Seyferts}
The X-ray spectra of Seyferts represent a superposition of multiple physical processes  in the galactic nucleus and surrounding gas. Phenomenologically the components of these spectra are measured to show a power-law continuum with a cutoff at $\sim$300 keV in the form of $E^{-\Gamma}\exp(-E/E_c)$, absorption from surrounding gas, emission lines, and a continuum hump, the called "reflection component", and a soft excess of emission at $\le2$keV, often approximated by a blackbody or a power-law.  According to the currently popular unification models, this primary continuum may be absorbed by circumnuclear material, with the degree of absorption related to the inclination of the symmetry axis of the accretion disk:  low-luminosity variants of such absorbed AGN are Seyfert II galaxies.  

Physically the primary continuum is thought to originate from multiple Compton scatterings of thermal disk photons in an optically thin (or at most moderately thick) hot corona above the disk \citep[see e.g.][]{kat76,poz77,sun80} with the high energy cutoff which roughly represents the temperature of the corona \cite[see e.g.][]{zdz94}.  The continuum slope (photon index) is determined by the Compton $y$-parameter which is a combination of the coronal temperature and optical depth. Reflection component appears as a result of the Compton reprocessed emission and bound-free absorption of the primary continuum by cold matter in the accretion disk and the surrounding gas \citep{lig88,mag95}.

As an aside, it is worth mentioning that X-ray binaries (XRBs) are also accretion disk systems, although the central black hole mass is of solar mass size. In fact, X-ray spectra of Seyferts resemble those of XRBs in hard state \citep{zdz99}. XRBs can extend this emission to MeV region with a power-law \citep{mcc94,gie99}.  Although MeV power-law tail has never been confirmed in Seyferts, some models predict the existence of such MeV power-law tail \cite[see e.g.][]{ino08}, in which thermal and non-thermal electrons coexist in the corona above the accretion disk. This scenario is naturally expected if the hot corona is heated by the magnetic reconnection \citep{liu02}. Non-thermal electrons are known to exist at Solar flares \citep[e.g.][]{shi95} and Earth's magnetotail \citep{lin05} where magnetic reconnection occurs. In the context of this model, with non-thermal component having $\sim4$\% of the total electron energy, MeV gamma-ray background can be explained by the same population of Seyferts that makes up the CXB as shown below \citep{ino08}. Observationally, the Oriented Scintillation Spectroscopy Experiment ({\it OSSE}) clearly detected emission up to 500 keV in the spectrum of the brightest Seyfert 1 NGC 4151 \citep{joh97}. Beyond $\sim200$ keV, the spectrum steepens. By combining the flux upper limit data above 500 keV, the maximum allowed non-thermal fraction is 15\% \citep{joh97}.

In this paper, we consider two intrinsic spectral models for Seyferts. One is thermal spectral model which has a power-law continuum with a cutoff \citep[see e.g.][]{ued03}. We adopt $\Gamma=1.9$ and $E_c=300$ keV.  The other is thermal plus non-thermal spectral model \citep[see][for details]{ino08}. We adopt the same parameters as in \citet{ino08}, but setting the thermal cutoff-energy at 300 keV. For the Compton reflection component, we use a Compton reflection model \citep{mag95} (developed for the XSPEC package as "pexrav"), assuming a solid angle of $2\pi$, an inclination angle of $\cos i = 0.5$, and solar abundance for all elements. To calculate absorbed spectra, we use an absorption model called "wabs" developed for the XSPEC package. 

\subsection{Blazars}
The multi-wavelength studies of blazars show that the overall spectra have two pronounced continuum components: one peaking between infrared and X-rays and another in the gamma-ray regime \citep{fos98,kub98}. The lower energy component is produced by synchrotron radiation, while the higher energy component is produced by the inverse Compton (IC) scattering of ambient seed photons by the same electrons \cite[see e.g.][]{ulr97,ghi98}. The target seed photon can be synchrotron radiation in the jet, in the synchrotron self-Compton (SSC) model \cite[see e.g.][]{jon74}, or external radiation such as emission from accretion disk, broad line region, or dusty torus, in the external radiation Compton (ERC) model \cite[see e.g.][]{der93,sik94}. 

Blazars can be classified into three subclasses by their spectra: high-energy peaked BL Lacertae objects (HBLs), low-energy peaked BL Lac objects (LBLs), and flat spectrum radio quasars (FSRQs).  The overall emission of HBLs may be explained by the SSC scenario, while that of FSRQs may be explained by the ERC scenario. X-ray spectra of HBLs, low-luminosity blazars,  show the softest spectra among them with photon index $\Gamma\sim2-3$, and this X-ray emission is the highest observable energy tail of the synchrotron component.  Since FSRQs   significantly contribute to the cosmic X-ray and MeV gamma-ray background as compared to BL Lacs \citep{aje09}, we focus on FSRQs only for blazars hereinafter.   X-ray spectra of FSRQs are harder and this emission is the lowest observable energy tail of the IC component. For FSRQs, we assume the blazar spectral energy distribution (SED) with an empirical double power-law model:
\begin{equation}
\frac{dN}{dE}\propto\left[\left(\frac{E}{E_b}\right)^{\Gamma_1}+\left(\frac{E}{E_b}\right)^{\Gamma_2}\right]^{-1},
\end{equation}
where we set $E_b=3$ MeV, $\Gamma_1=1.6$, and $\Gamma_2=2.9$ following \citet{aje09}. The average photon index of FSRQs observed by {\it Swift}-BAT at 15--55 keV is $1.6\pm0.3$ \citep{aje09}. Theoretically, FSRQs' spectra are expected to show a break and spectral softening at MeV band \citep[see e.g.][]{ino96}. However, $E_b$ and $\Gamma_2$ of FSRQs at MeV band are not constrained by observations due to the difficulty of the MeV gamma-ray measurement. The values of them here are artificially selected to explain the MeV background by FSRQs. If $E_b$ is at $\sim$MeV region, FSRQs can significantly contribute to the MeV background by choosing appropriate $\Gamma_2$. At GeV band, {\it Fermi} has observed 310 FSRQs whose mean value of photon index above 0.1 GeV is $2.42\pm0.17$ \citep{ack11_2FGL_AGN}. However, the photon index of the MeV gamma-ray background spectrum is $\sim$2.8 \citep{wat97}. This suggests $\Gamma_2\sim2.8$ to explain the MeV background by FSRQs. Therefore, if the MeV background is explained by FSRQs, MeV and GeV FSRQs may be different populations or FSRQs have a complex SED shape.

\section{X-ray Luminosity Function}
\label{sec:xlf}

To obtain the background radiation spectrum and angular power spectrum in the X-ray band, an XLF of sources is required. XLF gives the comoving number density at each luminosity and each redshift. We briefly review XLFs of Seyferts and blazars in this section.

\subsection{Seyferts}
Various X-ray surveys allowed to determine the evolution of Seyferts including unobscured and moderately obscured sources \citep[see e.g.][]{ued03,has05,gil07,air10}. X-ray photons above $\sim2$ keV are relatively unaffected by absorption for moderate column density ($N_{\rm H}\lesssim10^{23}\ \mathrm{cm^{-2}}$). XLF studies at the 2--10 keV band have revealed that luminosity-dependent density evolution (LDDE) models reproduce the observed XLFs at various redshift and luminosity ranges \citep{ued03,laf05,sil08,ebr09,yen09}. LDDE predicts that the shape of the XLF changes with redshift, with the faint-end slope flattening as redshift increases. This evolution is also characterized by a shift in the peak of the space density towards lower redshifts for lower luminosities, so-called downsizing. \citet{air10} suggested a more complex evolution model, luminosity and density evolution (LADE) model. LADE predicts a fixed shape of the XLF at all redshifts, but varies the normalization of the XLF.

In this study, we follow the \citet{ued03} LDDE XLF at 2-10 keV, since LDDE is confirmed to be adequate at 0.5--2 keV \citep[e.g.][]{miy00,has05} and 2--10 keV \citep[e.g.][]{ued03,laf05,sil08,ebr09,yen09} and  the distribution function of the neutral hydrogen column density is not available for the LADE model \citep{air10}. The comoving number density $\rho_X$ in the LDDE is given as:
\begin{equation}
\rho_X(L_X, z,N_{\rm H}) = \rho_X(L_X,0)f(L_X,z)\eta(N_{\rm H};L_X, z),
\end{equation}
where $L_X$ is the X-ray luminosity, $z$ is the redshift, and $N_{\rm H}$ is the neutral hydrogen column density. $\rho_X(L_X, 0)$ is the AGN XLF at present.  This is characterized by the faint-end slope index $\gamma_1 $, the bright-end slope index $\gamma_2$, and the break luminosity $L_X^*$, as:
\begin{equation}
\rho_X(L_X,0)=A_X \left[ \left( \frac{L_X}{L_X^*} 
\right)^{\gamma_1} + \left( \frac{L_X}{L_X^*} \right)^{\gamma_2} \right]^{-1} \ ,
\end{equation}
where $A_X$ is the normalization parameter having a dimension of volume$^{-1}$. 

The function $f(L_X,z)$ describes the density evolution, which is given by the following form:
\begin{eqnarray}  
  f(L_X,z)=\left\{\begin{array}{ll}
      (1+z)^{p_1} & z \le z_c(L_X), \\
      (1+z_c(L_X))^{p_1}
      \left( \frac{1+z}{1+z_c(L_X)} \right)^{p_2} & z > z_c(L_X), \\
    \end{array}\right.
\end{eqnarray} 
where $z_c$ is the redshift of evolutionary peak, given as
\begin{eqnarray}
z_c(L_X)=\left\{\begin{array}{ll}
    z_c^* & L_X \ge L_a, \\
    z_c^*(L_X/L_a)^\alpha & L_X < L_a. \\
    \end{array}\right.
\end{eqnarray}

The function $\eta(N_{\rm H};L_{\rm X},z)$ describes the distribution of absorption column density, which is given by the following form in the XLF \citep{ued03}:
\begin{eqnarray}  
\eta(N_{\rm H};L_{\rm X},z)
 &=\left\{\begin{array}{ll}
     2-\frac{5+2\epsilon}{1+\epsilon}\psi(L_{\rm X},z)&  (20.0 \leq \log N_{\rm H} < 20.5), \\
	\frac{1}{1+\epsilon}\psi(L_{\rm X},z) & (20.5 \leq \log N_{\rm H} < 23.0), \\
	\frac{\epsilon}{1+\epsilon}\psi(L_{\rm X},z) & (23.0 \leq \log N_{\rm H} < 24.0), 
    \end{array}\right.
\end{eqnarray} 
where  $\epsilon=$1.7 and 
\begin{equation}  
\psi(L_{\rm X},z)={\rm min}\{\psi_{\rm max}, {\rm max}[0.47-0.1(\log L_{\rm X} - 44.0), 0])\},
\end{equation}
for which
\begin{equation}
\psi_{\rm max}=\frac{1+\epsilon}{3+\epsilon}.
\end{equation}

The parameters obtained by the fit to the observed data of X-ray AGNs in \citet{ued03} are shown in Table \ref{XLF-parameters}. We set the minimum of the X-ray luminosity as $L_{X,\rm min}$=10$^{41.5}$ erg s$^{-1}$, the same as in \citet{ued03}.

As discussed in the introduction, an absorbed Seyfert population can contribute to the CXB significantly at $\gtrsim10$ keV. One of the main interests of X-ray AGN studies is the population of the Compton thick AGNs. The column density of the Compton thick AGNs is larger than $ N_{\rm H}=1/\sigma_T\simeq 1.5\times10^{24}\ {\rm cm^{-2}}$, where $\sigma_T$ is the Thomson scattering cross section. Here, we assume the fraction of the Compton thick AGNs between $24.0\le\log N_{\rm H}<25.0$ to be the same as that of the population at  $23.0\le\log N_{\rm H}<24.0$ following \citet{ued03} to explain the CXB by Seyferts. However, recent study of the {\it Swift}-BAT hard X-ray AGN samples revealed that the number density of the Compton thick AGNs is a factor of 3--4 less, at least locally, than that required to explain the CXB at hard X-ray \citep[see e.g.][]{tre09}. The Compton thick AGN population has not been fully resolved due to the necessity for imaging with hard X-ray instruments, and those, provided with, {\it NuSTAR} and {\it Astro-H} will resolve this population further beyond the local Universe. As shown below, the angular power spectrum at hard X-rays will be another probe to study the Compton thick AGN population.

Although the distribution function of absorption column density $\eta(N_{\rm H};L_{\rm X},z)$ is not available for the LADE model \citep{air10}, we can test the LADE model by assuming the same absorption column density distribution as in the LDDE. When we adopt the LADE model \citep{air10}, the overall normalization of the CXB needs to decrease by a factor of $\sim30$\% and no change is required to the spectral shape. Regarding the angular power spectrum, its shape does not change but the normalization will decrease by a factor of $\sim50$\% and $\sim10$\% at 0.5--2 keV and at 2--10 keV, respectively, by assuming the {\it e-Rosita} all sky survey sensitivity (see Figures. \ref{fig:Cl_0.5-2keV} and \ref{fig:Cl_2-10keV}). Our conclusion, which depends on the shape of the angular power spectrum, will not significantly change even if we adopt the LADE model for the Seyfert evolution.

\begin{deluxetable}{ccrrrrrrrrcrl}
\tabletypesize{\scriptsize}
\tablecaption{The parameters of the AGN XLF \label{XLF-parameters}}
\tablewidth{0pt}
\tablehead{
\colhead{} & \colhead{Ueda et al. 2003}  & \colhead{Ajello et al. 2009}\\
\colhead{} & \colhead{Seyfert}  & \colhead{FSRQ}\\
\colhead{} & \colhead{2-10 keV} & \colhead{15-55 keV} 
}
\startdata
$A_X$$^a$ & $50.4\pm3.3$  & $0.533\pm0.104$ \\
$\log_{10}$$L^*_X$$^b$ & $43.94_{-0.26}^{+0.21}$  & $44.0$ \\
$\gamma_1$ & $0.86\pm 0.15$  & - \\
$\gamma_2$ & $2.23\pm 0.13$  & $3.45\pm0.20$  \\
$z_c^*$ & $1.9$  & -\\
$\log_{10}$$L_a$$^b$ & $44.6$  & - \\
$\alpha$ & $0.335\pm 0.07$  & - \\
$p_1$ & $4.23\pm0.39$  & $3.72\pm0.50$ \\
$p_2$ & $-1.5$  & - \\
$p_3$ & -   & $-0.32\pm0.08$ \\
\enddata
\tablenotetext{a}{In units of $10^{-7} {\rm Mpc}^{-3}$.}  
\tablenotetext{b}{In units of ${\rm erg/s}$.}
\end{deluxetable}

\subsection{Blazars}
Gamma-ray studies of blazars indicated that LDDE well represents the evolution of EGRET blazars \citep{nar06,ino09}. The recent study of {\it Fermi} FSRQs confirmed that LDDE provides a good description of the evolution of FSRQs \citep{aje12_fermi}. However, in the X-ray band, the situation is different. Due to the beaming effect, the number density of blazars is less than Seyferts. Currently, {\it Swift}--BAT has done the deepest survey above 15 keV \citep{bau12}. \citet{aje09} studied the cosmological evolution of X-ray blazars using three years of data from {\it Swift}--BAT AGN survey with 26 FSRQs and 12 BL Lacs. The evolution of the FSRQs is reproduced by a PLE model. Since the number of samples for the \citet{aje09} XLF is limited, in the future study, it is necessary to use the XLF converted from other wavelength LFs such as radio or gamma-ray \citep[e.g.][]{aje12}, and by using the luminosity correlation \citep[see][for the case of radio galaxies]{ino11}. In this paper, we adopt this latest blazar XLF model \citep{aje09}.

The blazar XLF by \citet{aje09} is given in the form of pure luminosity evolution (PLE) model. PLE predicts the same shape of XLF at all redshifts, but shifts the shape with luminosity. The comoving number densities $\rho_X$ in the PLE is given as:
\begin{equation}
\rho_X(L_X, z) = \rho_X(L_X/e(z),0).
\end{equation}
The local XLF is characterized by a single power-law function with the slope index $\gamma_2$ and the break luminosity $L_X^*$, as:
\begin{equation}
\rho_X(L_X,0)=A_X \left( \frac{L_X}{L_X^*} \right)^{-\gamma_2}.
\end{equation}
The function $e(z)$ describes the evolution factor independent of luminosity, which is given by the following forms:
\begin{equation}
e(z)=(1+z)^{p_1 +p_3z}
\end{equation}  
The parameters obtained by the fit to the observed data of X-ray FSRQs in A09 are shown in Table \ref{XLF-parameters}. They also assume an evolving minimum luminosity as
\begin{equation}
L_\mathrm{X,Min}(z)=L_\mathrm{X,Min,0}\times e(z),
\end{equation}
where $L_\mathrm{X,Min,0}$ is the minimum luminosity as at $z=0$. We set $L_\mathrm{X,Min,0}=3\times10^{44} \ {\rm erg \ s^{-1}}$ \citep{aje09}.

\begin{figure}
\begin{center}
\centering
\plotone{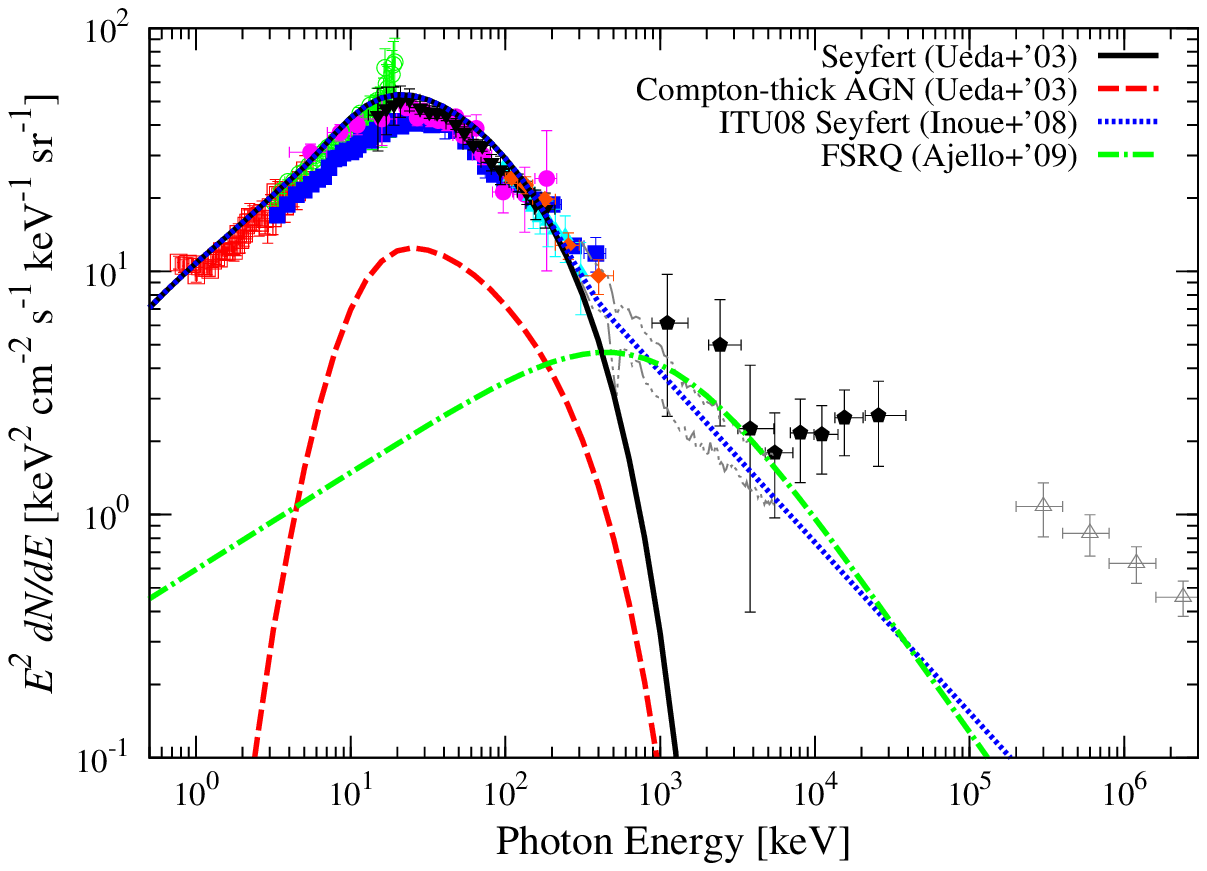} 
\caption{The cosmic X-ray and MeV gamma-ray background spectrum. Solid, dashed, dotted, and dot-dashed curve shows the contribution from Seyferts with $20<\log N_{\rm H}<25$ \citep{ued03}, Compton thick AGNs, i.e. Seyferts with $24<\log N_{\rm H}<25$ \citep{ued03}, Seyferts with non-thermal electrons in the coronae \citep{ino08}, and FSRQs \citep{aje09}. The cosmic X-ray background spectrum data of {\it ASCA} \citep[open squares,][]{gen95}, {\it RXTE} \citep[open circles,][]{rev05}, {\it HEAO}-1 A2 \citep[filled squares,][]{gru99}, {\it INTEGRAL} \citep[filled circles][]{chu07}, {\it HEAO}-1 A4 \citep[filled up-triangles,][]{kin97}, \textit{Swift}-BAT \citep[filled down-triangles,][]{aje08}, {\it SMM} \citep[triple dot-dashed,][]{wat97}, Nagoya--Ballon \citep[filled diamonds,][]{fuk75}, COMPTEL \citep[filled hexagons,][]{wei00}, and {\it Fermi} \citep[open triangles,][]{abd10_egrb} are shown in the figure. For the {\it SMM} data, the triple dot-dashed curve shows the 1-$\sigma$ uncertainty region. Although {\it Fermi} has measured the background spectrum up to 100 GeV \citep{abd10_egrb}, we plot their result up to 30 GeV to show the CXB and the MeV background clearly. \label{fig:CXB_spec}}
\end{center}
\end{figure}

\subsection{Cosmic X-ray and MeV Gamma-Ray Background Intensity}

 The unresolved background flux at the observed energy $E$ can be expressed as
 \begin{equation}
 I(E) = \int_{0}^{z_{\rm max}} dz \frac{d^2V}{dzd\Omega}\int_{L_{\rm min}}^{L(F_{\rm lim}(E), z)} dL \int_{N_{\rm H,min}}^{N_{\rm H,max}} dN_{\rm H}F(L,z,E)\rho(L,z,N_{\rm H}),
 \label{eq:CXB}
 \end{equation}
 where $F(L,z,E)$ is the observed photon flux from a source having a luminosity $L$ at redshift $z$ and $F_{\rm lim}(E)$ is the flux limit at the energy $E$.  In the case of blazars, we ignore the term of $N_{\rm H}$ in Eq. \ref{eq:CXB}. When we consider the total (resolved + unresolved) background flux\footnote{Although the background flux literally means the unresolved flux from the sky, the cosmic infrared, optical, and X-ray background flux usually mean the total flux which includes flux from resolved sources and unresolved sources \citep[e.g.][]{ino13,ued03,aje09}.}, we substitute $L(F_{\rm lim}(E), z)$ to $L_{\rm max}$, where we set $L_\mathrm{max}=10^{48} \ \mathrm{erg/s}$ for Seyferts \citep{ued03} and $L_\mathrm{max}=10^{50} \ \mathrm{erg/s}$ for blazars \citep{aje09}.

Fig. \ref{fig:CXB_spec} shows the contribution to the cosmic X-ray and MeV gamma-ray background spectra from Seyferts \citep{ued03}, Compton-thick AGNs  \citep{ued03}, Seyferts with non-thermal tails \citep{ino08}, and FSRQs \citep{aje09} using the spectral models in Section \ref{sec:sed} and the LFs in Section \ref{sec:xlf} together with observational data of {\it ASCA} \citep{gen95}, {\it RXTE} \citep{rev05}, {\it HEAO}-1 A2 \citep{gru99}, {\it INTEGRAL} \citep{chu07}, {\it HEAO}-1 A4 \citep{kin97}, \textit{Swift}-BAT \citep{aje08}, {\it SMM} \citep{wat97}, Nagoya--Ballon \citep{fuk75}, COMPTEL \citep{wei00}, and {\it Fermi} \citep{abd10_egrb}. By including the Compton-thick AGNs, we can adequately fit the CXB spectrum at 1--200 keV by the Seyfert population \citep{ued03}. Seyferts with non-thermal electrons in coronae \citep{ino08} and FSRQs \citep{aje09} can explain the MeV background.

\section{Cosmic X-ray and MeV Gamma-Ray Background Anisotropy}
In this section, we review the formalism to analytically calculate angular power spectra of cosmic background anisotropy \citep{peb80,lah97,and06}. Detailed formulation is given in Appendix \ref{sec:A_aps}.

The angular power spectrum of the CXB from point sources such as AGNs is given by
\begin{equation}
C_l = C_l^P + C_l^C,
\end{equation}
where the first term $C_l^P$ is the Poisson (shot noise) term and the second term $C_l^C$ is the correlation term \citep{peb80,and07_blz,and07_dm}. The shot noise term does not depend on the multipole $l$, while the correlation term reflects the intrinsic spatial correlation of sources. The multipole term $l$ is related to the angular separation $\theta$ in the sky as $l\simeq {180}/{\theta}$, where $\theta$ is the angular scale in the sky in units of degrees.

The two terms are related to the spatial power spectrum through
\begin{eqnarray}
\label{eq:ClP}
C_l^P &=& \int dz \frac{d^2V}{dzd\Omega}\int dL\int dN_{\rm H} F(L,z)^2\rho_X(L,z,N_{\rm H})\\  \nonumber
C_l^C&=&\int dz \frac{d^2V}{dzd\Omega}P_{\rm lin}(k=\frac{l}{r(z)},z)\\
&&\times \left[\int dL\int dN_{\rm H} b_{\rm AGN}(L,z)F(L,z)\rho_X(L,z,N_{\rm H})\right]^2,
\label{eq:ClC}
\end{eqnarray}
where $P_{\rm lin}(k,z)$ is the power spectrum of the linear matter density fluctuation as a function of the wave number $k=l/r$, $r(z)$ is the proper distance, and $b_{\rm AGN}(L,z)$ is the bias factor of AGNs against dark matter. We use the linear transfer function given in \citet{eis99} for $P_{\rm lin}(k,z)$. The integration range is the same as Eq. \ref{eq:CXB} for the unresolved background flux. We also assume the Limber approximation which means that fluctuation does not change strongly and which is valid for small angular separation, $l\gtrsim6$ corresponding to $\theta\lesssim30^\circ$. In the case of blazars, we ignore the term of $N_{\rm H}$ in Eqs. \ref{eq:ClP} and \ref{eq:ClC}.

The 1-$\sigma$ statistical error in the measurement of the angular power spectrum is given by 
\begin{equation}
(\delta C_l)^2 = \frac{2C_l^2}{(2l+1)\Delta l f_{\rm sky}},
\label{eq:Cl_err}
\end{equation}
where $\Delta l$ is the bin size in the multipole space and $f_{\rm sky}$ is a fraction of the sky covered by observations \citep[see][for details]{and07_blz,and07_dm}. Hereinafter we assume the all sky survey, with $f_{\rm sky}=1$ (such as {\it e-Rosita}) and set $\Delta l=0.5l$. Eq. \ref{eq:Cl_err} shows that the statistical error is reduced by removing as many point sources as possible or by measuring as high $l$ (small $\theta$) as possible.

\subsection{Bias Factor of Active Galactic Nuclei}

The bias factor of AGNs is a key to understanding  the environment  of the AGN formation in the cosmic history. The bias factor represents the clustering strength of AGNs compared with dark matter (See Eq. \ref{eq:A_bias}). Clustering of AGNs has been studied with large samples in the optical large survey, such as the Two-degree Field Quasar Redshift Survey \citep[][]{cro05,por06} and the Sloan Digital Sky Survey \citep[SDSS][]{li06,ros09,she09}. The bias evolves from $b_{\rm AGN}\sim1.4$ at $z=0.5$ \citep{ros09},  $b_{\rm AGN}\sim3$ at $z=2.2$ \citep{ros09}, to $b_{\rm AGN}\sim10$ at $z=4.0$ \citep{she09}. In X-rays, many papers have explored the angular clustering of AGNs  \citep{vik95,aky00,yan03,bas04,mul04_bias,gan06,puc06,car07,miy07,pli08,ebr09,bra11,ely12,kou13}. However, the bias parameter of AGNs has not been determined to agree neither between optical and X-ray nor amongst various X-ray studies. While both the angular and 3D correlation function of the X-ray bright AGNs detected by the {\it ROSAT} suggested that close to unity with the median redshift $z=0.4$ \citep{vik95,mul04_bias}, those from {\it Chandra} and {\it XMM-Newton} suggested stronger clustering \citep{yan03,bas05,gan06}. The inferred bias parameter from {\it XMM-Newton} Large Scale Structure survey \citep{gan06} is $\sim3.7$ at the median redshift $z=0.7$ \citep{and07_dm}.  The most recent X-ray study based on 1466 X-ray AGN samples at $0<z<3$ suggested $b_{\rm AGN}=2.26$ at the redshift $z=0.976$ \citep{kou13}. They also showed that the bias of X-ray AGNs is significantly higher than those of optically selected AGNs at each redshift. In this study, although it is known that $b_{\rm AGN}>1$ \citep{kou13}, we conservatively set $b_{\rm AGN}=1$ independent of for redshift and luminosity for the simplicity, unless noted otherwise.

\section{Results}
\subsection{0.5-2 keV and 2-10 keV band}

\begin{figure*}
\begin{center}
\centering
\plotone{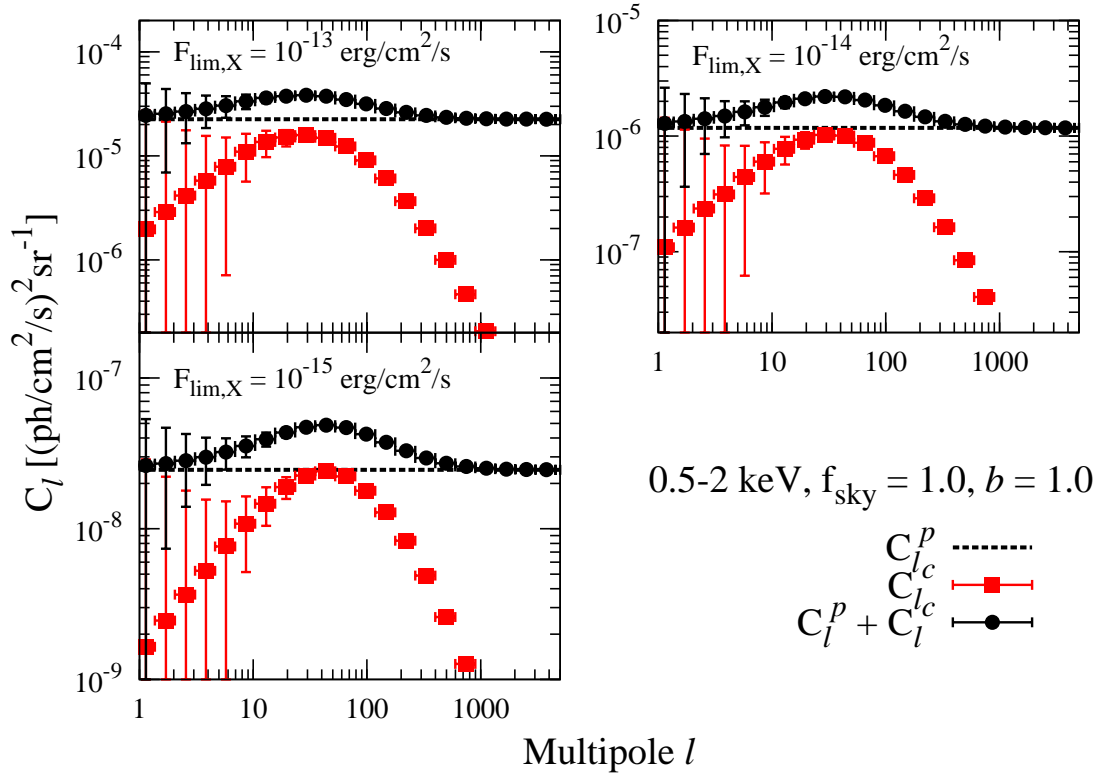} 
\caption{Predicted angular power spectra of Seyferts at 0.5-2 keV with $b=1.0$ following the \citet{ued03} XLF. Each panel shows the all sky survey case with the sensitivity limit shown in the panel. Filled circle and filled square points show the total angular power spectrum ($C_l^P+C_l^C$) and the correlation term $C_l^C$, respectively. The horizontal dashed line represents the Poisson (Shot noise) term $C_l^P$. The error bars show the 1$\sigma$ errors with $\Delta l=0.5l$. The scale of {\it y}-axis of each panel is different. 0.5-2 keV corresponds the soft band of {\it e-Rosita} is 0.5--2 keV and its sensitivity limit with a 4-year survey is $F_\mathrm{lim}\simeq10^{-14} \ {\rm erg/cm^2/s}$ at this band \citep{mer12}. \label{fig:Cl_0.5-2keV}}
\end{center}
\end{figure*}

Each panel of fig. \ref{fig:Cl_0.5-2keV} shows the results for the angular power spectra of Seyferts at 0.5--2 keV for different sensitivity limit with 1-$\sigma$ error bars. We adopt $b_{\rm AGN}=1$ and $f_{\rm sky}=1$ here. We find that it would be possible to measure the correlation term of Seyferts with the sensitivity of $10^{-13} \ {\rm erg/cm^2/s}$ or better. At the large multipole region $l\gtrsim500$ (corresponding to $\sim 22$ arcmin), the deviation of total angular power spectrum from the poisson term is hardly seen. As fainter point sources are resolved, the Poisson term is reduced and the correlation term will be more clearly detected.

\begin{figure*}
\begin{center}
\centering
\plotone{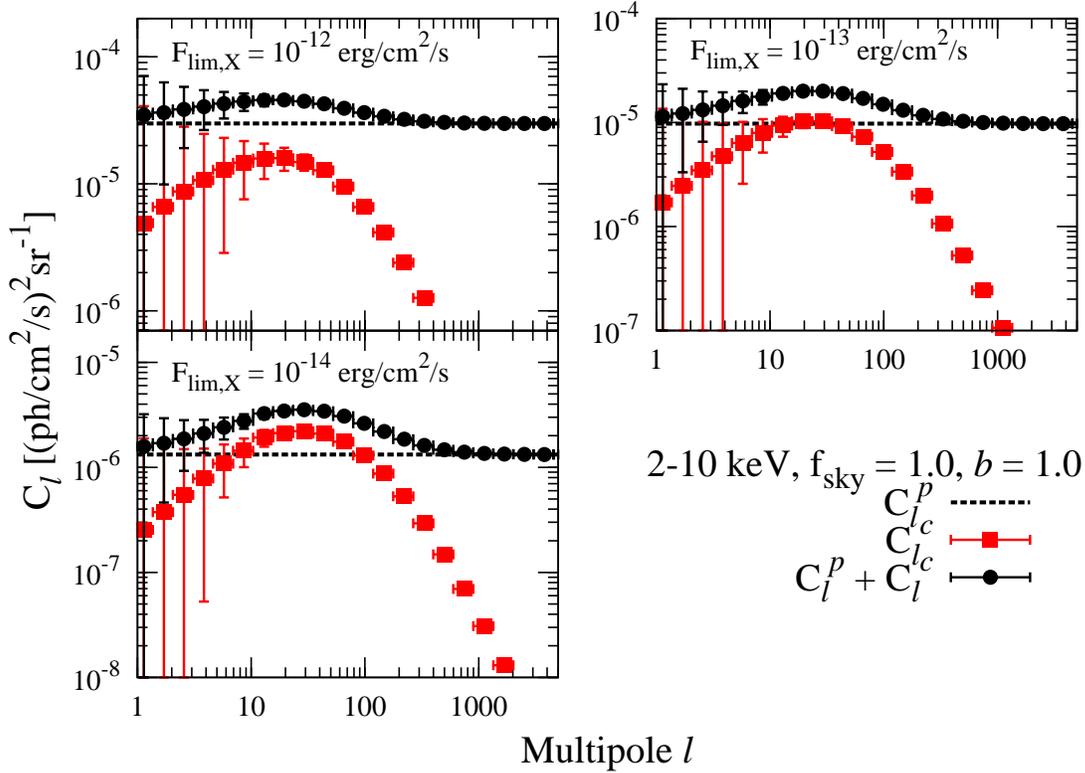} 
\caption{Same as Fig. \ref{fig:Cl_0.5-2keV}, but for 2-10 keV and different sensitivity limits as indicated in each panel. The hard band of {\it e-Rosita} is 2--10 keV and its sensitivity limit with a 4-year survey is $F_\mathrm{lim}\simeq10^{-13} \ {\rm erg/cm^2/s}$ at this band \citep{mer12}\label{fig:Cl_2-10keV}}
\end{center}
\end{figure*}

The angular power spectra of Seyferts at 2--10 keV are shown in Fig. \ref{fig:Cl_2-10keV} for various sensitivity limits. We can measure the correlation term of Seyferts with the sensitivity of $10^{-12} \ {\rm erg/cm^2/s}$ or better. Similar to the case of 0.5--2 keV, at the large multipole region $l\gtrsim500$, the deviation of total angular power spectrum from the poisson term is hardly seen.

{\it e-Rosita} will perform an all sky survey with the sensitivity of $10^{-14}\ {\rm erg/cm^2/s}$  at soft band (corresponding to 0.5-2 keV) and $10^{-13}\ {\rm erg/cm^2/s}$ at hard band (corresponding to 2-10 keV) with a 4-year survey \citep{mer12}. The point spread function of {\it e-Rosita} is $\sim28$ arcsec (corresponding to $l\sim2.3\times10^4$) at 1 keV for the survey mode. {\it e-Rosita} will clearly detect the angular power spectrum of CXB and its correlation signal around $10\lesssim l\lesssim1000$ at both of 0.5-2 keV and 2-10 keV even with $b_{\rm AGN}=1$, if the CXB at these energy bands is composed of Seyferts. Since the Poisson term does not depend on the multipole $l$, we can derive the Poisson term using $C_l$ at $l\gtrsim500$.

\begin{figure}
\begin{center}
\centering
\plotone{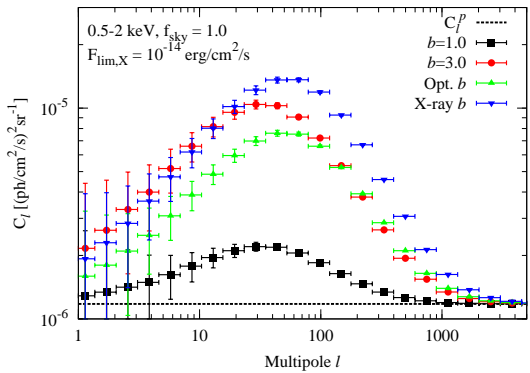} 
\caption{Predicted total (Poisson + Correlation) angular power spectra at 0.5-2 keV following the \citet{ued03} XLF. Different bias parameters. We set $F_{{\rm lim},X}=10^{-14} \ {\rm erg/cm^2/s}$ to demonstrate the capability of {\it e-Rosita} at 0.5-2 keV. All sky survey mode is assumed $f_{\rm sky}=1$. Square, circle, upper-triangle, and lower-triangle point shows the total angular power spectrum ($C_l^P+C_l^C$) with $b_{\rm AGN}=1$, 3, $b_{\rm AGN}$ following the optical evolution \citep{kou13}, and $b_{\rm AGN}$ following the X-ray evolution \citep{kou13} respectively.  The horizontal dashed line represents the Poisson (Shot noise) term $C_l^P$. The error bars show the 1$\sigma$ errors with $\Delta l=0.5l$. 
\label{fig:Cl_0.5-2keV_bias}}
\end{center}
\end{figure}

\begin{figure}
\begin{center}
\centering
\plotone{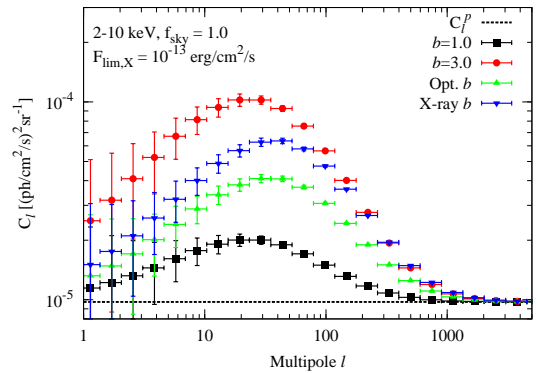} 
\caption{Same as Fig. \ref{fig:Cl_0.5-2keV_bias}, but for 2-10 keV.  We set $F_{{\rm lim},X}=10^{-13} \ {\rm erg/cm^2/s}$ to demonstrate the capability of {\it e-Rosita} at 2-10 keV. 
\label{fig:Cl_2-10keV_bias}}
\end{center}
\end{figure}

Figs. \ref{fig:Cl_0.5-2keV_bias} and \ref{fig:Cl_2-10keV_bias} show the total (Poisson + correlation) angular power spectra from Seyferts at 0.5-2 keV for the sensitivity limit of $10^{-14}\ {\rm erg/cm^2/s}$ and at 2-10 keV for the sensitivity limit of $10^{-13}\ {\rm erg/cm^2/s}$, respectively, to demonstrate the capability of {\it e-Rosita}.  Four different bias models are considered. We plot here the cases with constant bias $b_{\rm AGN}=1$, $b_{\rm AGN}=3$, the evolving bias parameters inferred from optically selected quasars, and that from X-ray selected AGNs \citep[see the right panel of Fig. 8 of ][for details]{kou13}. At both energy bands, angular power spectra of CXB enable us to clearly distinguish these models by {\it e-Rosita}. 

We do not need to divide the angular power spectrum into the Poisson term and the correlation term to derive the bias parameter. The bias will be derived by using the total angular power spectrum which is the directly observable value. Once we obtain the XLF of Seyferts from X-ray deep survey studies \citep[e.g.][]{ued03,has05,gil07}, the angular power spectrum of CXB obtained by the future X-ray all sky survey will enable us to verify what kind of the bias evolution model is favored. In particular, {\it e-Rosita}, which covers both energy bands, can evaluate the bias parameter at different energies.

\subsection{10-30 keV band}
 
\begin{figure*}
\begin{center}
\centering
\plotone{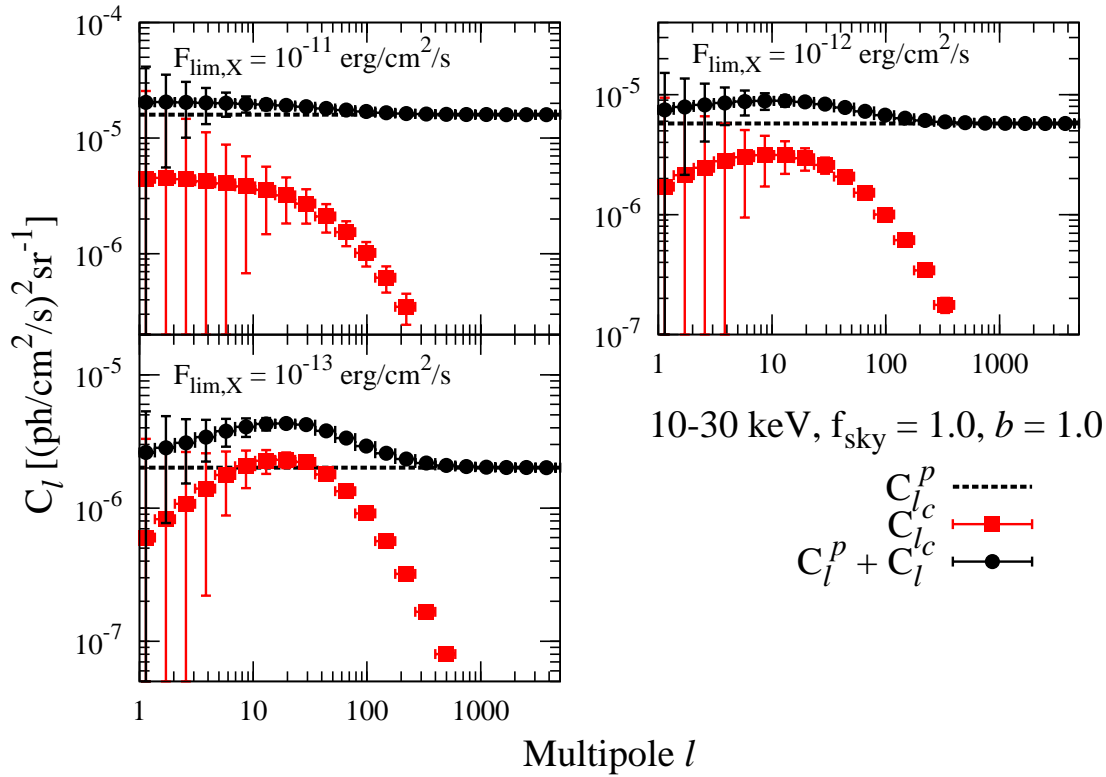} 
\caption{Same as Fig. \ref{fig:Cl_2-10keV}, but for 10--30 keV. \label{fig:Cl_10-30keV}}
\end{center}
\end{figure*}

\begin{figure*}
\begin{center}
\centering
\plotone{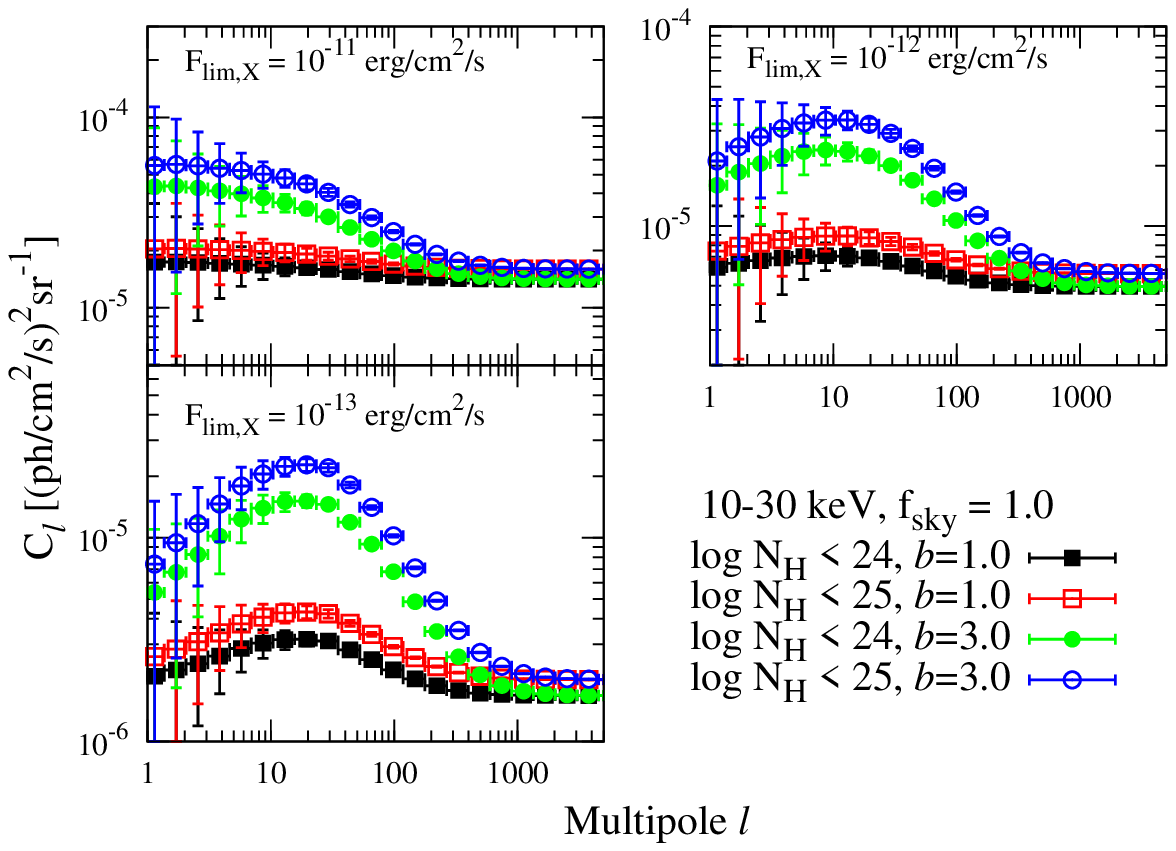} 
\caption{Predicted angular power spectra of Seyferts  for various $b$ and $\log N_{\rm H}$ at 10--30 keV following the \citet{ued03} XLF. Each panel shows the all sky survey case $f_{\rm sky}=1$ with the sensitivity limit shown in the panel. Filled and open points show Seyferts with $\log N_{\rm H}<24$ and $\log N_{\rm H}<25$, respectively. Square and circle point shows the case with $b_{\rm AGN}=1$ and $b_{\rm AGN}=3$, respectively. The error bars show the 1$\sigma$ errors with $\Delta l=0.5l$. The scale of {\it y}-axis of each panel is different. 
 \label{fig:Cl_10-30keV_NH}}
\end{center}
\end{figure*}

Fig. \ref{fig:Cl_10-30keV} shows the angular power spectra of Seyferts by all sky survey observations, analogous to Fig. \ref{fig:Cl_2-10keV} but for 10--30 keV band. The deviation of the correlation term from the Poisson term is difficult to be measured with the sensitivity limit of $10^{-11}\ {\rm erg/cm^2/s}$ due to the statistical errors. We need the sensitivity better than $10^{-12}\ {\rm erg/cm^2/s}$ to detect the correlation term at hard X-ray band. Since the current most sensitive all sky hard X-ray survey is done by {\it Swift}-BAT with the sensitivity level of $\sim10^{-11}\ {\rm erg/cm^2/s}$ \citep{bau12}, one order of magnitude more sensitive instruments are required to measure the correlation term. At 30--50 keV, we obtained similar results.

Angular power spectrum at hard X-ray will be another probe to study the Compton thick AGN population. Fig. \ref{fig:Cl_10-30keV_NH} shows the total (Poisson + correlation) angular power spectra of Seyferts for bias $b_{\rm AGN}=1$ and 3 and absorption column density with $\log N_{\rm H}<24.0$ and $\log N_{\rm H}<25.0$. Each panel represents the case with the sensitivity limit shown in the panel. Low statistical errors are crucial to determine the fraction of the Compton thick AGNs. Hence, point sources should be removed as many as possible to reduce the statistical errors. In the case of $F_{{\rm lim}, X}=10^{-11}\ {\rm erg/cm^2/s}$, it is difficult to distinguish the contribution of Compton thick AGNs with $b_{\rm AGN}=1$ due to large statistical errors, while it may be possible to see the difference at $l\gtrsim20$ with $b_{\rm AGN}=3$. If we can achieve the sensitivity of $F_{{\rm lim}, X}=10^{-12}\ {\rm erg/cm^2/s}$ or better, we can distinguish the contribution of Compton thick AGNs even with $b_{\rm AGN}=1$.

Can pointing observatories measure angular power spectra of the background radiation? Although the sensitivity limit of hard X-ray all sky survey is still above the required sensitivity for the angular power spectrum study, the pointing observatories such as {\it NuSTAR} and {\it Astro-H} can achieve the sensitivity of $\sim10^{-14} \ \mathrm{erg/cm^2/s}$ at 10 keV for 100 ks observations \citep{har10,tak12}. As an example, the field of view of {\it NuSTAR} is 13 arcmin. If {\it Nustar} can do one hundred 100 ks pointing observations in the extragalactic sky during its operation, $f_\mathrm{sky}$ will be $10^{-4}$. Following Eq. \ref{eq:Cl_err}, the statistical error will be two orders of magnitude more enhanced than the case of $f_\mathrm{sky}=1$. This large statistical error makes difficult to measure the angular power spectrum of the background radiation with such a small field of view instruments.

\subsection{MeV band}

Measurement of MeV gamma-rays is difficult. The dominant process in a detector is Compton scattering and huge background of photons are produced in the MeV instruments themselves. COMPTEL onboard the {\it CGRO} satellite is the only instrument that observed the entire MeV sky and it discovered only $\sim$30 gamma-ray sources at 0.75--30 MeV \citep{sch00_comptel}. Thus, the MeV sky has not been fully investigated. {\it Astro-H} which is scheduled to be launched in 2015 will have a sub-MeV instrument, soft gamma-ray detector \citep[SGD;][]{taj10}. The SGD covers a wide energy range from 40 keV up to 600 keV \citep{tak12}. The field of view (FOV) of the SGD varies with energy. A BGO collimator defines $\sim$10 deg FOV at high energies, while a fine collimator restricts the FOV to $\sim$0.6 deg below $\sim$150 keV.  Even though Astro-H is designed to perform pointing observations, the 10 deg FOV of the SGD above 150 keV will allow for a wide sky coverage over the course of the ASTRO-H mission, which is essential to study the MeV background. A number of future projects are currently proposed to observe the MeV sky such as CAST \citep{nak12}, DUAL \citep{bal12}, GRIPS \citep{gre12} and SMILE \citep{tak11}. Recently, various ballon experiments have been carried out to test the performance of instruments \citep{tak11,ban11}. All of these future instruments will resolve the MeV sky in the coming decades. 

\begin{figure*}
\begin{center}
\centering
\plotone{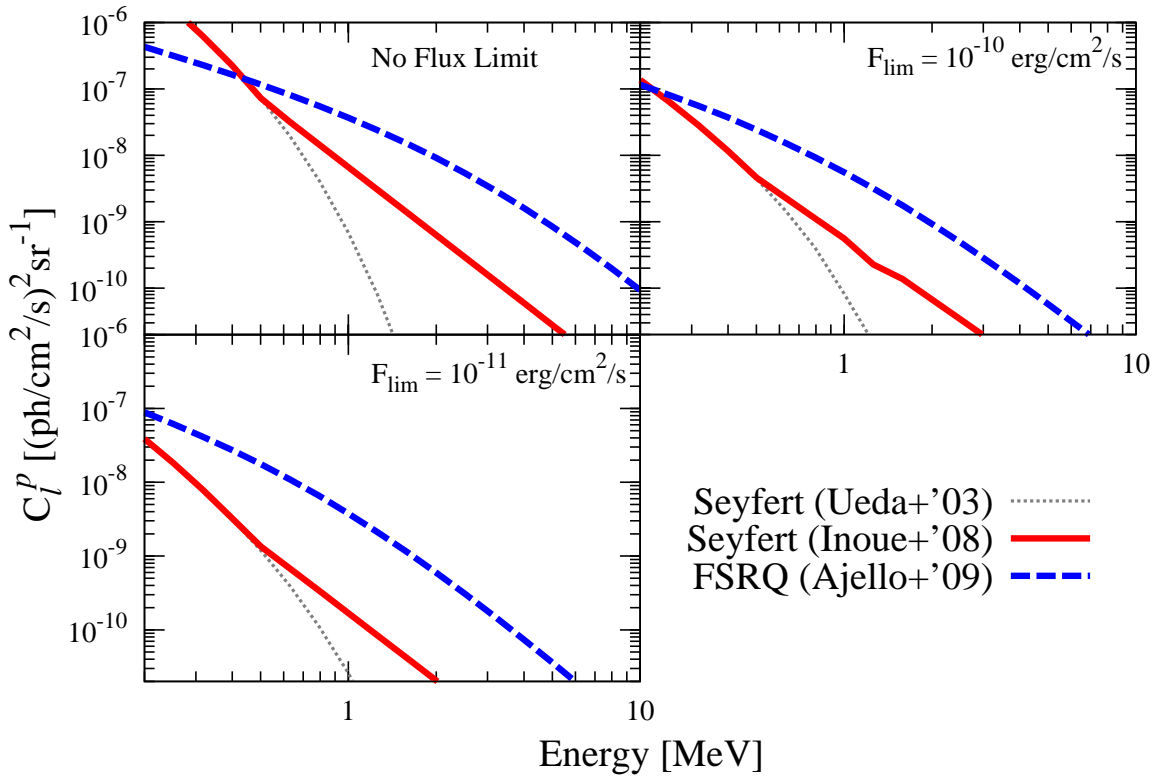} 
\caption{Predicted poisson term of the angular power spectrum of the cosmic MeV background at 200 keV-- 10 MeV. Each panel shows the all sky survey case $f_{\rm sky}=1$ with the $\nu F_\nu$ sensitivity limit shown in the panel. Solid and dashed curve corresponds to Seyferts with non-thermal electrons in coronae \citep{ino08} and FSRQs \citep{aje09}, respectively, assuming the MeV background is explained by them. For reference, we also plot the model of Seyferts with thermal cutoff \citep{ued03} by dotted curve which does not explain the MeV background radiation.  \label{fig:Cl_MeV}}
\end{center}
\end{figure*}

Even with those instruments, it is not as easy to resolve the MeV sky as in soft X-ray \citep[see e.g.][]{bau04}. However, one can distinguish the origin of the MeV background by measuring its angular power spectrum. Fig. \ref{fig:Cl_MeV} shows the Poisson term of the angular power spectra of Seyferts with non-thermal components in coronae \citep{ino08} and FSRQs \citep{aje09} with various $\nu F_\nu$ sensitivity limit. For reference, we also plot Seyferts with simple thermal cutoff spectra \citep{ued03}, but note that those do not explain the MeV background. Since the contribution of the correlation term is negligible in this energy region and the assumed flux limits, the angular power spectrum is dominated by the Poisson term. This Poisson term measurement is useful enough to distinguish the origin of the MeV background. We do not show the expected errors which are highly dependent on the range of observed multipoles. Errors can be estimated from Equation \ref{eq:Cl_err}. By using high multipole value and wide multipole bin size, the errors will become small. For example, if we select $l=100$ and $\delta l=100$ ($50\le l\le150$), the expected uncertainty will be $\delta C_l \sim 0.01 C_l$.

Left-top panel of Fig. \ref{fig:Cl_MeV} shows the case in which no sources are resolved. We integrate Eq. \ref{eq:ClP} between $L_{\rm min}$ and $L_{\rm max}$. Even if the MeV sky is not resolved into point sources, we can distinguish the origin of the cosmic MeV background. The difference of the $C_l^p$ of Seyferts \citep{ino08} and FSRQs \citep{aje09} is more than an order of magnitude. The reason why we can clearly distinguish the origin is as follows. Seyferts are fainter but more numerous than blazars. These two differences are able to make future MeV instruments clearly detect the origin of the MeV gamma-ray sky through the angular power spectrum of the sky (see Eq. \ref{eq:ClP}). Therefore, ballon experiments may be able to distinguish the origin of the MeV background sky, although it may suffer from little photon statistics. As \citet{wei00} put an upper limit on the relative deviations from isotropy of the MeV background, it will be worth revisiting the COMPTEL data in future studies.

In the process of resolving the sources contributing to the MeV background via improvements of sensitivity and angular resolution, the contribution of Seyferts to the angular power spectrum at sub-MeV region decreases more rapidly than FSRQs. This is because Seyferts dominate the sub-MeV gamma-ray background at $\lesssim400$ keV (see Fig. \ref{fig:CXB_spec}). With the sensitivity of $10^{-10} \ \mathrm{erg/cm^2/s}$ in $\nu F_\nu$ close to the sensitivity limit of the COMPTEL \citep{sch00}, we can clearly distinguish the Seyfert scenario \citep{ino08} vs. the FSRQ scenario \citep{aje09}. Future MeV sky survey instruments will easily distinguish the origin of the MeV background. However, we note that there will be a very significant background from an instrument itself in the case of the Compton camera technique. Since it may contribute to the angular power spectrum at some level, it is crucial to reduce background events as many as possible. The SGD on board {\it Astro-H} is expected to reduce such background significantly \citep{tak12}.

\section{Discussion and Conclusions}
In this paper we have studied the angular power spectra of Seyferts and blazars from 0.5 keV - 10 MeV. We have shown that {\it e-Rosita} can detect the spatial clustering of Seyferts including the bias information at 0.5--2 keV and 2--10 keV, which is a long standing problem in optical and X-ray AGN survey studies. As the {\it e-Rosita} AGN sample also allows us to study the bias information \citep{kol13},  it will be complementary to each other.

In order to distinguish the population of Compton-thick AGNs, which is believed to be relevant at hard X-ray band (10-30 keV and 30-50 keV), we need to detect the correlation term. However, the sensitivity better than $10^{-12}\ {\rm erg/cm^2/s}$ is required for this purpose, so the present, best all-sky survey by Swift-BAT \citep[with $\sim10^{-11}\ {\rm erg/cm^2/s}$][]{bau12} is insufficient. Future improvement of the hard X-ray survey instruments is necessary for this study.

At MeV band, we can clearly distinguish the origin of the MeV background candidates, Seyferts \citep{ino08} and FSRQs \citep{aje09} even with current MeV instruments including ballon experiments. However, this requires that these missions can measure the angular power spectrum of the sky. Future MeV instruments such as SGD onboard {\it Astro-H}, DUAL, GRIPS, and SMILE will easily disentangle the origin of the MeV background via covering a wide solid angle of the sky with their expected sensitivities. 

If the origin of the MeV background is non-thermal emission from Seyfert \citep{ino08}, this implies that magnetic reconnection heats the corona above the disk and accelerate non-thermal electrons in the corona. As discussed in \citet{ino08}, this scenario will be also tested by future observations of individual sources. For example, the expected flux from NGC 4151, which is the brightest Seyfert galaxy in the hard X-ray sky \citep{saz07}, is $\sim3\times10^{-5}(E/{\rm MeV})^{-0.8}\ {\rm MeV \ cm^{-2}\ s^{-1}}$ \citep{ino08}, which can not be detected by COMPTEL but by the future MeV instruments such as CAST, DUAL, GRIPS, and SMILE. If it is FSRQs \citep{aje09}, this implies that there are two distinct FSRQ populations in MeV and GeV because of the spectral difference between MeV and GeV background. This will suggest that there are two different populations in FSRQs between MeV and GeV. This may pose a problem to the AGN unification scheme \citep{urr95}. Therefore, probing the MeV background is another handle on to understanding AGN physics. 

In our study, we use the power spectrum of linear dark matter fluctuation \citep{eis99}. At a small angular separation, however, effects of non-linear dark matter fluctuation on the correlation term can not be ignored \citep{sel00}. Non-linear contributions will become important at the scale of $\lesssim1.5 h^{-1}\ \mathrm{Mpc}$ \citep{kou13}. This corresponds to $l\gtrsim\sim 600$ at $z\sim0.3$. $\sim50$\% of the unresolved CXB flux comes from inside of $z=0.3$ at the sensitivity of {\it e-Rosita} at soft and hard band. Since the correlation term has a peak at $l\sim100$, the non-linear effect will not change our results significantly.

The required multipole scale for the study of angular power spectrum at X-ray bands is $l\lesssim500$. This corresponds to $\theta\gtrsim22$ arcmin. The point spread function of {\it e-Rosita} is $\sim28$ arc sec at 1 keV for the survey mode. Therefore, {\it e-Rosita} will clearly detect the angular power spectrum. On the other hand, MeV instruments do not have as small point spread function as X-ray instruments have. However, the Poisson term which is the key to understanding the origin of the MeV background does not depend on the multipole. 

Gravitational lensing by clusters of galaxies may change the angular power spectrum, since the lensed sources are strongly clustering and are amplified around the lensing cluster. It has been argued that the observed AGN luminosity function could be significantly affected by lensing \citep[e.g.][]{tur80,tur84}. The fraction of lensed AGNs at $z\lesssim4.3$ is expected to be less than 2\% at the SDSS limiting sensitivity \citep{wyt02}. Therefore, the lensing will not affect our results significantly. 

Other populations such as galaxies are responsible for a fraction of CXB, although it is expected to be $\sim2$\% \citep{per03,bau04}. They may also alter the shape of the angular power spectrum. As galaxies are fainter and more numerous than AGNs, their Poisson term will be weaker than that of AGNs and their correlation term contribution arises at different multipole due to the difference of the distribution in the sky. The lensing may also alter fluctuation signatures of CXB and the MeV background, if they are dominated by galaxies.  A deficit of surface brightness within the central regions of massive galaxy clusters, which is a strong lensing cluster, after removing detected sources has recently measured with the Herschel Space Observatory. The amplitude of the deficit is the same as the full intensity of the lensed cosmic infrared background radiation which is dominated by galaxies \citep{zem13}. 

We thank Kazuo Hiroi, Megumi Shidatsu, Takeshi Tsuru, Tatsuya Sawano, Atsushi Takada, and Toru Tanimori for useful comments. YI acknowledges support by the Research Fellowship of the Japan Society for the Promotion of Science (JSPS). YI thanks the hospitality of the Center for Cosmology and AstroParticle Physics (CCAPP) at the Ohio State University. KM is supported by NASA through Hubble Fellowship grant No. 51310.01 awarded by the Space Telescope Science Institute, which is operated by the Association of Universities for Research in Astronomy, Inc., for NASA, under contract NAS 5-26555.

\appendix
\section{Angular Power Spectrum of Cosmic Background Radiation}
\label{sec:A_aps}

 Following Eq. \ref{eq:CXB}, the total CXB background intensity received from the direction $ \hat{\pmb{r}}$ can be expressed as
 \begin{eqnarray}
 I(\hat{\pmb{r}} ,E) &=& \int_{0}^{z_{\rm max}} dz \frac{d^2V}{dzd\Omega}\int_{L_{\rm min}}^{L(F_{\rm lim}, z)} dL F(L,z,E)\rho(L,z; \hat{\pmb{r}}), \label{eq:A_cxb}\\
 &=&\frac{c}{4\pi}\int_{0}^{z_{\rm max}} dz \left|\frac{dt}{dz}\right|\int_{L_{\rm min}}^{L(F_{\rm lim}, z)}  dL L(E,z)\rho(L,z; \hat{\pmb{r}})\\
 &=& \frac{1}{4\pi}\int_{0}^{r(z_{\rm max})} dr  \int_{L_{\rm min}}^{L(F_{\rm lim}, z)}  dL L(E,z)\rho(L,z; \hat{\pmb{r}})
 \end{eqnarray}
 where we assume that the distribution in $L$ is statistically independent of position and $r$ is a proper distance corresponding to a redshift $z$. The integration term for the column density $N_{\rm H}$ is added to calculate the background flux from Seyferts. Hereinafter, we also do not show the term of $E$ and the integration range explicitly. The deviation of the CXB intensity from its mean value is
\begin{equation}
\delta I(\hat{\pmb{r}})  \equiv  I(\hat{\pmb{r}})  -  <I>. 
\end{equation}

Following \citet{peb80}, the autocorrelation function of the CXB for point sources is
\begin{eqnarray}
C(\theta)&=&<\delta I(\hat{\pmb{r}}_1)\delta I(\hat{\pmb{r}}_2)>\\ 
&=& <I(\hat{\pmb{r}}_1)I(\hat{\pmb{r}}_2)> - <I>^2,\\ \label{eq:A_Ctheta}
&=&\frac{1}{16\pi^2}\int dr_1\int dr_2 \xi(\pmb{r}_1-\pmb{r}_2)\left[\int dL_1 L_1(z_1) \rho_X(L_1,z_1; \hat{\pmb{r}}_1)\right]\left[\int dL_2 L_2(z_2) \rho_X(L_2,z_2; \hat{\pmb{r}}_2)\right],
\end{eqnarray}
where $\theta$ is the angle between $\hat{\pmb{r}}_1$ and $\hat{\pmb{r}}_2$ in units of radians and $\xi(\pmb{r})$ is the two-point correlation function of AGNs, which gives the excess probability for finding a neighbor at $\pmb{r}$. We set $\pmb{r}_1=r_1\hat{\pmb{r}}_1$ and $\pmb{r}_2=r_2\hat{\pmb{r}}_2$.

The Poisson term of the angular power spectrum is obtained by setting $\theta=0$ for Eq. \ref{eq:A_Ctheta} \citep[see \S. 58 of][]{peb80}
\begin{eqnarray}
C_l^p =\int dz \frac{d^2V}{dzd\Omega}\int dL F(L,z)^2\rho_X(L,z).
\end{eqnarray}

The correlation term of the angular power spectrum of the CXB is related to the correlation function by setting $\theta\neq0$ as
\begin{equation}
C_l^C=\int_{\theta\neq 0}d^2\theta e^{-i\pmb{l}\cdot\pmb{\theta}}C(\theta).
\label{eq:A_cor}
\end{equation}

For the simplicity, we use small separation approximation, so-called the Limber approximation. Eq. \ref{eq:A_Ctheta} becomes
\begin{eqnarray}
C(\theta)&=&\frac{1}{16\pi^2}\int dr\int du\xi(u\hat{\pmb{r}}+r(z)\theta\hat{\pmb{\theta}},z)\left[\int dL L \rho_X(L,z)\right]^2,\\
&=&\int dz \frac{d^2V}{dzd\Omega}\int du\frac{\xi(u\hat{\pmb{r}}+r(z)\theta\hat{\pmb{\theta}},z)}{16\pi^2(1+z)^2r(z)^2}\left[\int dL L \rho_X(L,z)\right]^2,
\end{eqnarray}
where $r$ is $(r_1+r_2)/2$, $u$ is $r_2-r_1$, and we use $dr_1dr_2=drdu$.

Then, Eq. \ref{eq:A_cor} becomes
\begin{eqnarray}
C_l^C&=&\int d^2\theta \int dz \int du\frac{d^2V}{dzd\Omega}e^{-i\pmb{l}\cdot\pmb{\theta}}\frac{\xi(u\hat{\pmb{r}}+r(z)\theta\hat{\pmb{\theta}},z)}{16\pi^2(1+z)^2r(z)^2}\left[\int dL L \rho_X(L,z)\right]^2\\
&=&\int d^2\theta \int dz \int du\int \frac{d^3k}{(2\pi)^3}\frac{d^2V}{dzd\Omega}e^{-i\pmb{l}\cdot\pmb{\theta}}\frac{P_{\rm AGN}(k,z)e^{i\pmb{k}\cdot(u\hat{\pmb{r}}+r\theta\hat{\pmb{\theta}})}}{16\pi^2(1+z)^2r(z)^2}\left[\int dL L \rho_X(L,z)\right]^2\\
&=&\int d^2\theta \int dz \int du\int \frac{dk_{\parallel}d^2k_{\perp}}{(2\pi)^3}\frac{d^2V}{dzd\Omega}\frac{P_{\rm AGN}(k,z)e^{-ik_{\parallel}\cdot u}e^{i\pmb{\theta}\cdot(r\pmb{k}_{\perp}-\pmb{l})}}{16\pi^2(1+z)^2r(z)^2}\left[\int dL L \rho_X(L,z)\right]^2\\
&=&\int dz \int dk_{\parallel}d^2k_{\perp}\frac{d^2V}{dzd\Omega}\frac{P_{\rm AGN}(k,z)\delta_{D}(k_{\parallel})\delta^2_D(r\pmb{k}_{\perp}-\pmb{l})}{16\pi^2(1+z)^2r(z)^2}\left[\int dL L \rho_X(L,z)\right]^2\\
&=&\int dz \frac{d^2V}{dzd\Omega}\frac{P_{\rm AGN}(k=l/r,z)}{16\pi^2(1+z)^2r(z)^4}\left[\int dL L \rho_X(L,z)\right]^2\\
&=&\int dz \frac{d^2V}{dzd\Omega}P_{\rm AGN}(k=\frac{l}{r},z)\left[\int dL F(L,z)\rho_X(L,z)\right]^2,
\label{eq:A_ang_agn}
\end{eqnarray}
where we decomposed the wave number $\pmb{k}$ by the components parallel and perpendicular to $\pmb{r}$, $\pmb{k} = \pmb{k}_{\parallel}+\pmb{k}_{\perp}$, and used $d^3k=dk_{\parallel}d^2k_{\perp}$. We also used the relation $d_L(z)=(1+z)r(z)$, flux-luminosity relation, and the following Fourier transformation
\begin{equation}
\xi(u\hat{\pmb{r}}+r(z)\theta\hat{\pmb{\theta}},z) = \int \frac{d^3k}{(2\pi)^3}P_{\rm AGN}(k,z)e^{i\pmb{k}\cdot(u\hat{\pmb{r}}+r\theta\hat{\pmb{\theta}})}.
\end{equation}

We also need to consider the bias of AGNs against dark matter. The power spectrum of AGNs is given by
\begin{equation}
P_{\rm AGN}(r,z;L_1,L_2)=b_{\rm AGN}(L_1,z)b_{\rm AGN}(L_2,z)P_{\rm lin}(r,z),
\label{eq:A_bias}
\end{equation}
where $b_{\rm AGN}$ represents the clustering strength of AGNs compared with dark matter and $P_{\rm lin}$ represents the power spectrum of linear dark matter density fluctuations. We use the linear transfer function given in \citet{eis99} to calculate $P_{\rm lin}(r)$.

Then, Eq. \ref{eq:A_ang_agn} becomes
\begin{equation}
C_l^C=\int dz \frac{d^2V}{dzd\Omega}P_{\rm lin}(k=\frac{l}{r},z)\left[\int dL b_{\rm AGN}(L,z)F(L,z)\rho_X(L,z)\right]^2.
\end{equation}
This is the same as Eq. \ref{eq:ClC}.

\end{document}